\documentclass[english,twocolumn]{revtex4}
\usepackage[T1]{fontenc}
\usepackage[latin9]{inputenc}
\usepackage[active]{srcltx}
\usepackage{amsmath}
\usepackage{graphicx}
\usepackage{esint}

\makeatletter
\@ifundefined{textcolor}{}
{%
 \definecolor{BLACK}{gray}{0}
 \definecolor{WHITE}{gray}{1}
 \definecolor{RED}{rgb}{1,0,0}
 \definecolor{GREEN}{rgb}{0,1,0}
 \definecolor{BLUE}{rgb}{0,0,1}
 \definecolor{CYAN}{cmyk}{1,0,0,0}
 \definecolor{MAGENTA}{cmyk}{0,1,0,0}
 \definecolor{YELLOW}{cmyk}{0,0,1,0}
}

\usepackage{dcolumn}% Align table columns on decimal point
\usepackage{bm}% bold math

\makeatother

\usepackage{babel}
\begin{document}

\title{A model for shock wave chaos}

\author{Aslan Kasimov }

\email{aslan.kasimov@kaust.edu.sa}

\author{Luiz Faria}

\address{King Abdullah University of Science and Technology (KAUST) \\
Division of Mathematical and Computer Sciences and Engineering\\
Thuwal 23955-6900, Saudi Arabia }

\author{Rodolfo R. Rosales}

\address{Massachusetts Institute of Technology\\
Department of Mathematics\\
Cambridge, MA, USA}
\begin{abstract}
We propose the following model equation: 
\[
u_{t}+\frac{1}{2}\left(u^{2}-uu_{s}\right)_{x}=f\left(x,u_{s}\right),
\]
that predicts chaotic shock waves. It is given on the half-line $x<0$
and the shock is located at $x=0$ for any $t\ge0$. Here $u_{s}\left(t\right)$
is the shock state and the source term $f$ is assumed to satisfy
certain integrability constraints as explained in the main text. We
demonstrate that this simple equation reproduces many of the properties
of detonations in gaseous mixtures, which one finds by solving the
reactive Euler equations: existence of steady traveling-wave solutions
and their instability, a cascade of period-doubling bifurcations,
onset of chaos, and shock formation in the reaction zone. 
\end{abstract}

\pacs{02.30.Jr, 47.10.ab, 47.40.Rs, 05.45.-a, 47.40.Nm, 47.70.Fw}

\maketitle

Shock waves arise in a wide range of physical phenomena: gas dynamics,
shallow-water flows, supernovae, stellar winds, traffic flows, quantum
fluids, and many others. The dynamics of shock waves can be quite
intricate and difficult to analyze due to the difficult nature of
the hyperbolic conservation laws that govern their evolution. The
theory of shock waves has a rich history beginning with the fundamental
contributions by Riemann in the middle of the 19th century. Nevertheless,
numerous open questions remain regarding the shock dynamics, especially
when classical gas-dynamical shock waves interact with additional
physical or chemical phenomena, such as magnetic and gravitational
fields, chemical reactions, radiation, and others. 

Our focus in this Letter is on one fascinating feature of a shock
wave propagating in a chemically active medium, namely \emph{shock-wave
chaos}. This is a phenomenon wherein the shock propagates with its
speed oscillating chaotically about a certain average. It has previously
been demonstrated to occur in gaseous \emph{detonations,} by solving
the reactive Euler equations \cite{Ng2005,HenrickAslamPowers2006}.
Detonations are shock waves in reactive mixtures that are sustained
by the chemical energy release in the mixture; the reactions, in turn,
are triggered and sustained by the heating provided by the shock compression. 

Analytical and numerical difficulties associated with solving the
reactive Euler equations motivated the introduction of simple analog
models, with the aim of capturing the essential nature of observed
detonation shocks. As it is well-known, Burgers \cite{burgers1948mathematical}
introduced his equation, $u_{t}+uu_{x}=u_{xx}$ (where the subscripts
$t$ and $x$ indicate partial derivatives), now a hallmark of hyperbolic
differential equations and shock wave theory, in the hope of capturing
the essential nature of turbulence with a simple and tractable model.
Following a similar idea, Fickett \cite{Fickett:1979ys,Fickett1985}
and shortly after him Majda \cite{Majda:1980zr}, introduced simple
analog models for detonations in the hope of gaining some insight
into the complicated behavior of detonation waves, as observed in
experiment and numerical simulations. Fickett's model is a modification
of the Burgers' equation, which introduces the effects of chemical
energy release. It takes the form: 
\begin{alignat}{1}
u_{t}+\frac{1}{2}(u^{2}+q\lambda)_{x}=0,\quad\lambda_{t}=\omega(\lambda,u),\label{eq: Fickett's Model}
\end{alignat}
where $u$ is the primary unknown mimicking density, temperature,
or pressure, $\omega$ is a rate function, and $q$ is a constant
playing the role of a chemical energy release. The chemistry here
is represented by an irreversible reaction $reactants\rightarrow products$,
with $\lambda$ being a normalized concentration of reaction products.
At the shock, $\lambda=0$ and $\lambda$ increases through the reaction
zone to reach $\lambda=1$ in the products.

Fickett's model has been shown to reproduce some of the features of
detonations \cite{Fickett:1979ys,fickett1985stability,Fickett1985},
most notably the steady-state structures. Still, the key unstable
character of detonations had not been reproduced within this model
until Radulescu and Tang \cite{Radulescu:2011fk} extended it to a
two-step chemistry with an inert induction zone followed by an energy-releasing
reaction zone. In \cite{Radulescu:2011fk}, the authors were able
to reproduce, with their analog model, the complexity of chaotic detonations
in the Euler equations. 

In this Letter, we propose a model consisting of a \emph{single} equation
that predicts steady traveling wave solutions, instability through
a Hopf bifurcation, and a sequence of period-doubling bifurcations
with subsequent chaotic dynamics. The onset of chaos in our equation
appears to follow the same scenario as in the logistic map \cite{may1976simple}.
For the reactive Euler equations and the Fickett's analog model, the
same scenario has been found \cite{HenrickAslamPowers2006,Radulescu:2011fk}.
We note that even though our model is still an analog, it is close
to the weakly nonlinear model of Rosales and Majda \cite{RosalesMajda:1983ly},
which is rationally derived from the Euler equations rather than postulated
as the Fickett's (or Majda's) models. However, our emphasis here is
not on the precise relationship of our model to the Euler equations,
but rather on presenting what we believe is the simplest partial differential
equation that is capable of capturing much of the richness of detonations
in the reactive Euler equations.

Our model is the following partial differential equation: 
\begin{equation}
u_{t}+\frac{1}{2}\left(u^{2}-uu_{s}\right)_{x}=f\left(x,u_{s}\right),\label{eq:KFR-equation}
\end{equation}
for $x<0$ and $t\ge0$, with an appropriate initial condition, $u\left(x,0\right)$.
Here $u_{s}\left(t\right)=u\left(0,t\right)$ is the boundary value
of the solution, which is not prescribed but follows by solving (\ref{eq:KFR-equation}),
as explained below. The source term $f$ needs to satisfy certain
integrability conditions, as also explained further below. 

Equation (\ref{eq:KFR-equation}) is a simple model for the reaction
zone of a detonation moving into a uniform state, in coordinates attached
to the leading shock. This can be seen by application of the Rankine-Hugoniot
shock conditions (see, e.g. \cite{CourantFriedrichs}) at $x=0$,
for (\ref{eq:KFR-equation}) extended by taking $f=0$ and $u=0$
for $x>0$. Indeed, the shock condition 
\begin{equation}
-V\left[u\right]+\frac{1}{2}\left[u^{2}\right]-\frac{1}{2}u_{s}\left[u\right]=0,
\end{equation}
where $V$ is the shock speed and $[z]=z^{+}-z^{-}$ is the jump of
$z$ across the shock (so that $\left[u\right]=-u_{s}$ and $\left[u^{2}\right]=-u_{s}^{2}$),
yields $V=0$. We assume that the shock satisfies the usual Lax entropy
conditions \cite{leveque1994numerical}, so that the characteristics
from both sides of the shock converge on the shock. That is, 
\begin{alignat}{1}
 & \frac{dx}{dt}|_{0-}=\left(u-\frac{u_{s}}{2}\right)_{x=0-}=\frac{u_{s}}{2}>0\quad\mbox{and }\\
 & \frac{dx}{dt}|_{0+}=\left(u-\frac{u_{s}}{2}\right)_{x=0+}=-\frac{u_{s}}{2}<0,
\end{alignat}
which require that $u_{s}>0$. Therefore, no boundary condition at
$x=0$ is necessary. Finally, we remark that $u_{s}$ is a measure
of the shock strength, since $\left[u\right]=-u_{s}$, and for that
reason we will analyze $u_{s}\left(t\right)$ in what follows when
describing the shock dynamics.

The most unusual feature of (\ref{eq:KFR-equation}) is that the equation
contains in it the boundary value of the unknown, $u_{s}\left(t\right)$.
This is in fact the key reason for the observed complexity of the
solutions and has a simple physical interpretation: the boundary information
from $x=0$ is propagated \emph{instantaneously} throughout the solution
domain, $x<0$, while there is a finite-speed influence propagating
from the reaction zone back toward the shock along the characteristics
of (\ref{eq:KFR-equation}). Importantly, in the Euler equations,
this situation occurs in a weakly nonlinear reactive shock wave where
the flow behind the shock is nearly sonic relative to the shock \cite{RosalesMajda:1983ly}.
One family of acoustic characteristics is then nearly parallel to
the shock, representing the slow part of the wave moving toward the
shock. The second family moves away from the shock and represents
the influence of the shock on the whole post-shock flow. This occurs
on a much faster time scale than the information flow toward the shock.
Our model makes this fast influence instantaneous. 

One can easily obtain the steady-state solution $u_{0}\left(x\right)$
of (\ref{eq:KFR-equation}) by solving 
\begin{equation}
\frac{1}{2}\left(u_{0}^{2}-u_{0}u_{0s}\right)'=f\left(x,u_{0s}\right),
\end{equation}
where the prime denotes the derivative with respect to $x$ and the
subscript $s$ denotes the shock state. The solution is 
\begin{equation}
u_{0}\left(x\right)=\frac{u_{0s}}{2}+\sqrt{\frac{u_{0s}^{2}}{4}+2\int_{0}^{x}f\left(y,u_{0s}\right)dy}.\label{eq:u0(x)}
\end{equation}
The choice of the steady-state shock strength 
\begin{equation}
u_{0s}=2\sqrt{2\int_{-\infty}^{0}f\left(y,u_{0s}\right)dy}\label{eq:u0s}
\end{equation}
corresponds to the Chapman-Jouguet speed in detonation theory \cite{FickettDavis79},
since then the characteristic speed at $x=-\infty$ is $u_{0}\left(-\infty\right)-u_{0s}/2=0$
indicating that the sonic point is reached only at an infinite distance
from the shock. This situation is analogous to the commonly used simple-depletion
kinetics in gaseous detonations \cite{FickettDavis79}. 

If one substitutes (\ref{eq:u0s}) into (\ref{eq:u0(x)}), the result
is 
\begin{equation}
u_{0}\left(x\right)=\frac{u_{0s}}{2}+\sqrt{2\int_{-\infty}^{x}f\left(y,u_{0s}\right)dy}.\label{eq:u0(x)_cj}
\end{equation}
Clearly, for the solution $u_{0}\left(x\right)$ to be real and bounded,
one must require that 
\begin{equation}
0\le\int_{-\infty}^{x}f\left(y,u_{0s}\right)dy<\infty
\end{equation}
for any $-\infty<x\le0$. This is the constraint on $f$ that we mentioned
earlier in the Letter. There may be other or additional, more stringent
conditions on $f$ in other circumstances, but their discussion is
outside the scope of this Letter.

Now we explore the fully nonlinear and unsteady solutions of (\ref{eq:KFR-equation}),
for the particular case 
\begin{equation}
f=\frac{q}{2}\frac{1}{\sqrt{4\pi\beta}}\exp\left[-\frac{\left(x-x_{f}\left(u_{s}\right)\right)^{2}}{4\beta}\right].\label{eq:f_gaussian}
\end{equation}
The function $f$ peaks at $x_{f}$, chosen here as $x_{f}=-k\left(u_{0s}/u_{s}\right)^{\alpha},$
where $k>0$ and $\alpha\ge0$ are parameters. We first rescale the
variables as follows: $u$ by $u_{0s}$, so that the dimensionless
steady-state shock strength is $1$, length by $l=k$, and time by
$\tau=l/u_{0s}$. From (\ref{eq:u0(x)_cj}), putting in all the dimensionless
variables and rescaling $\beta$ by $l^{2}$, we obtain (keeping the
same notation for the dimensionless variables and parameters) 
\begin{equation}
u_{0}\left(x\right)=\frac{1}{2}\left[1+\sqrt{\frac{1+\mathrm{erf}\left(\left(x+1\right)/2\sqrt{\beta}\right)}{1+\mathrm{erf}\left(1/2\sqrt{\beta}\right)}}\right],
\end{equation}
where $\mathrm{erf}\left(x\right)$ is the error function. The dimensionless
form of (\ref{eq:KFR-equation}) is 
\begin{alignat}{1}
 & u_{t}+\frac{1}{2}\left(u^{2}-uu_{s}\right)_{x}=a\exp\left[-\frac{\left(x+u_{s}^{-\alpha}\right)^{2}}{4\beta}\right],\label{eq:KFR-dimless}
\end{alignat}
where $a=1/\left[4\sqrt{4\pi\beta}\left(1+\mathrm{erf}\left(1/2\sqrt{\beta}\right)\right)\right]$.
Equation (\ref{eq:KFR-dimless}) contains only two parameters now,
$\alpha$ reflecting the shock-state sensitivity of the source function
$f$ (an analog of the activation energy in detonations) and $\beta$
reflecting the width of $f$ (an analog of the ratio between the reaction-zone
length and the induction-zone length).

\begin{figure}
\noindent \begin{raggedright}
\includegraphics[width=3.8in]{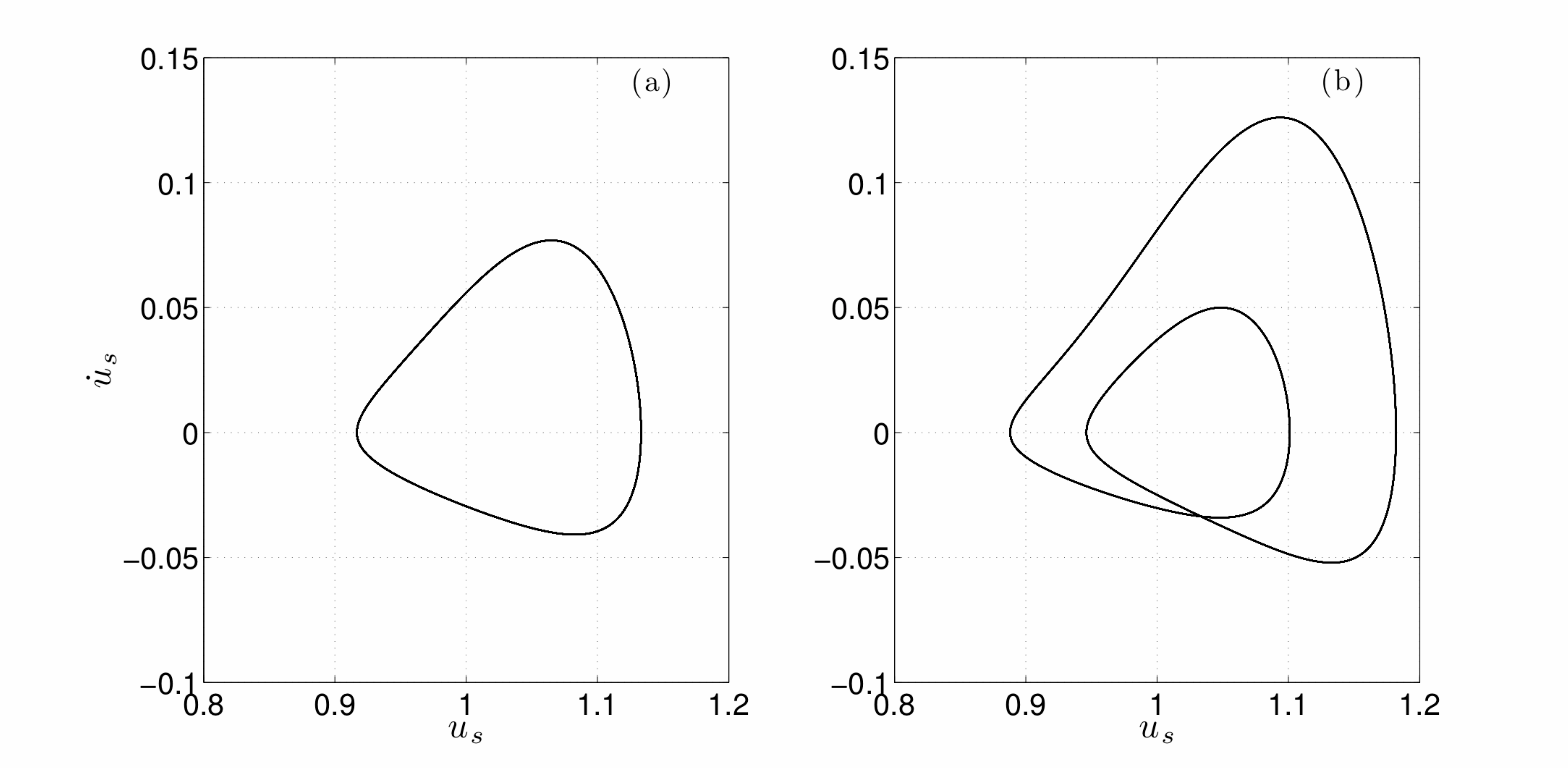}
\par\end{raggedright}

\caption{\label{fig:us(t)}The limit cycles in the plane of the shock strength
$u_{s}\left(t\right)$ vs $\dot{u}_{s}$ at $\alpha=4.70$ (a) and
$\alpha=4.85$ (b).}
\end{figure}

In the computations below, we use the shock-fitting algorithm of \cite{HenrickAslamPowers2006}
on a domain of length $L=10$ with $N=3000$ uniformly spaced grid
points. We fix $\beta=0.1$ in all calculations and vary $\alpha$
to capture instability and bifurcations. When long-time data, such
as the local maxima of $u_{s}\left(t\right)$ are needed, we compute
until $t=6000$. Simulations start with the steady-state solution
perturbed by numerical noise. In Fig. \ref{fig:us(t)} one can see
that a period doubling occurs as $\alpha$ is increased from $\alpha_{1}=4.70$
to $\alpha=4.85$. Below the critical value $\alpha_{c}\approx4.04$,
the steady solution is found to be stable. 

\begin{figure}
\noindent \begin{raggedright}
\includegraphics[width=3.7in]{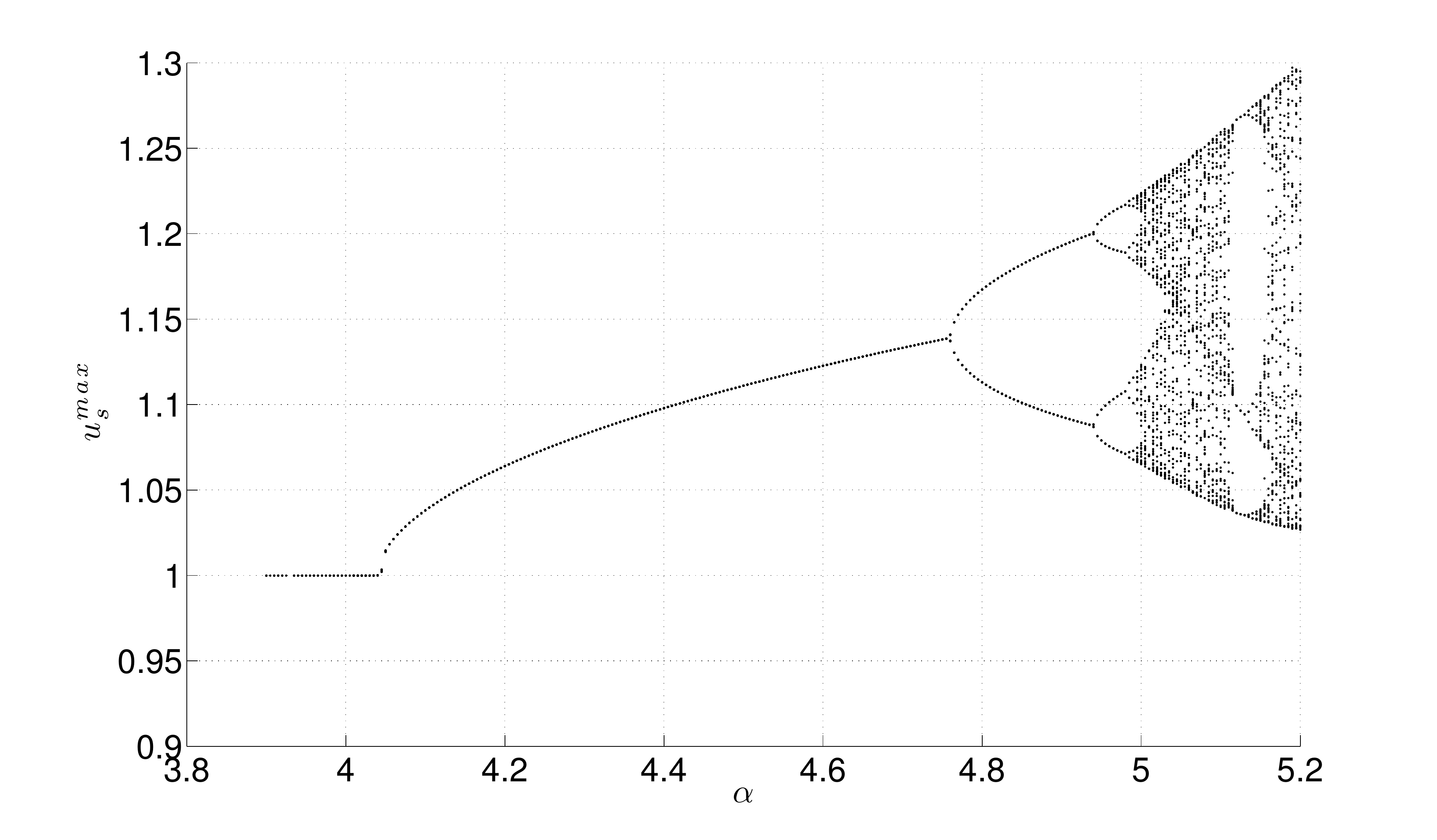}
\par\end{raggedright}

\caption{\label{fig:Bifurcation-diagram}Values of the local maxima $u_{s}^{max}$
of the shock strength as a function of the parameter $\alpha$; $u_{s}^{max}$
are calculated at sufficiently large $t$ to make sure the solution
has settled to its attractor. }
\end{figure}

If $\alpha$ is increased to large values, we observe that the reaction
zone extends significantly initially, but subsequently shrinks. Importantly,
as the reaction zone shrinks, another shock is formed within the reaction
zone which then overtakes the lead shock at $x=0$, exactly analogous
to what happens in the reactive Euler equations. Under these conditions,
the dynamics is no longer smooth and must be analyzed differently.
Therefore, we focus on moderately large $\alpha$, namely $\alpha<5.2$
for our particular choice of $\beta$, so that the dynamics is unstable,
but no internal shock waves appear to form. Remarkably, as $\alpha$
is increased, we observe a sequence of period-doubling bifurcations
that leads to chaotic solutions at $\alpha$ close to or slightly
larger than $5$, as seen in Fig. \ref{fig:Bifurcation-diagram}.
The onset of chaos apparently follows the same scenario as in the
logistic map \cite{may1976simple,strogatz1994nonlinear}. The bifurcation
diagram in Fig. \ref{fig:Bifurcation-diagram} was computed by solving
(\ref{eq:KFR-dimless}) until $t=6000$ for the range of $\alpha$
from $3.9$ to $5.2$, with an increment of $0.005$. For each $\alpha$,
we find the maxima of $u_{s}\left(t\right)$ between $t=5000$ and
$t=6000$ and plot them on the figure. Based on a sequence of three
period doublings, we estimated the Feigenbaum constant $\delta$ \cite{feigenbaum1983universal}
to be about $4.5$. This is in rough agreement with the well-known
value of $\delta=4.669...$ for the logistic map as well as that found
for detonations \cite{strogatz1994nonlinear,Ng2005,HenrickAslamPowers2006,Radulescu:2011fk}. 

\begin{figure}
\noindent \begin{raggedright}
\includegraphics[width=3.7in]{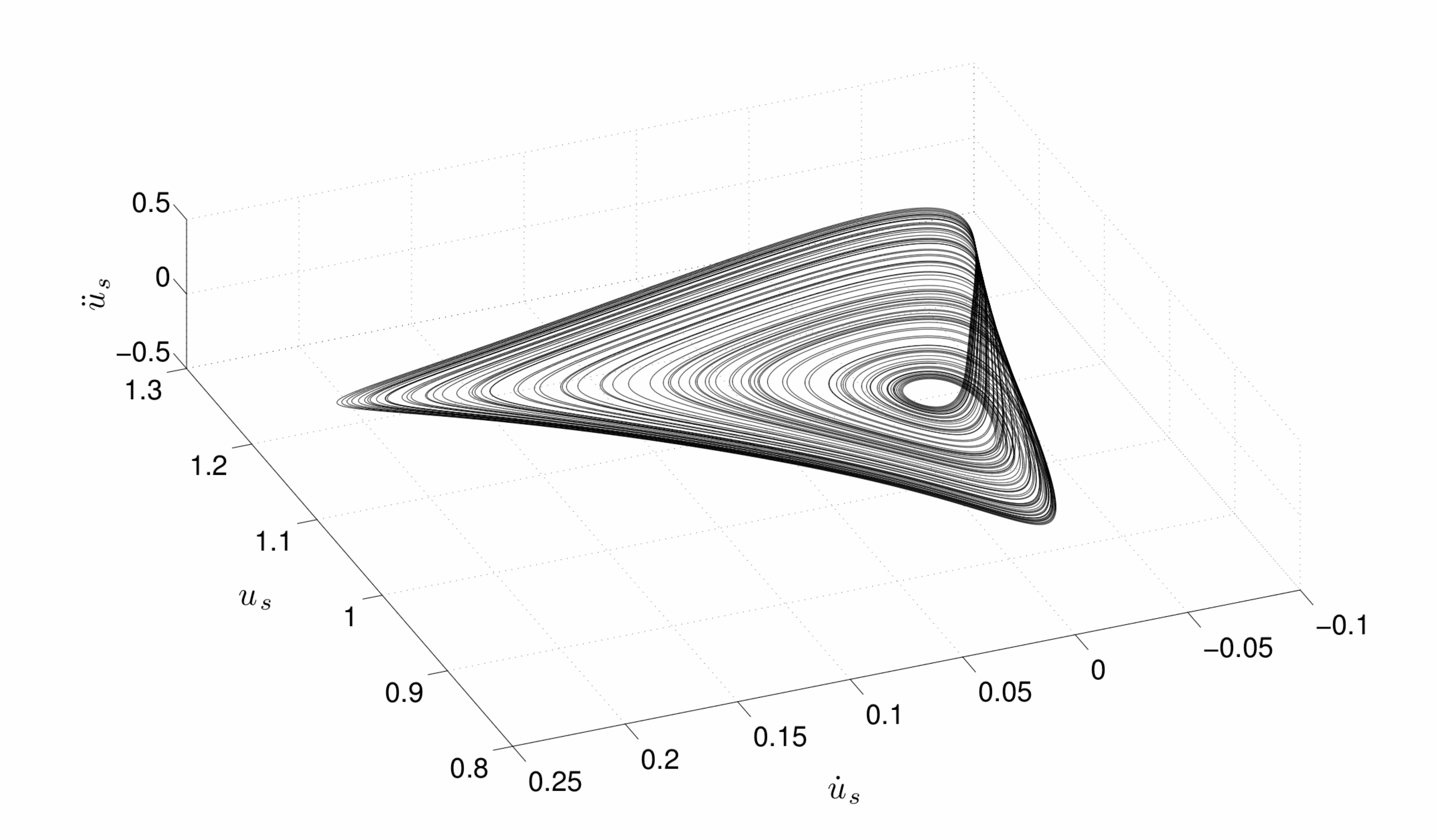}
\par\end{raggedright}

\caption{\label{fig:Phase-plane}The chaotic attractor in the space of $u_{s}$,
$\dot{u}_{s}$, and $\ddot{u}_{s}$ at $\alpha=5.1$.}
\end{figure}
\begin{figure}
\noindent \begin{raggedright}
\includegraphics[width=3.8in]{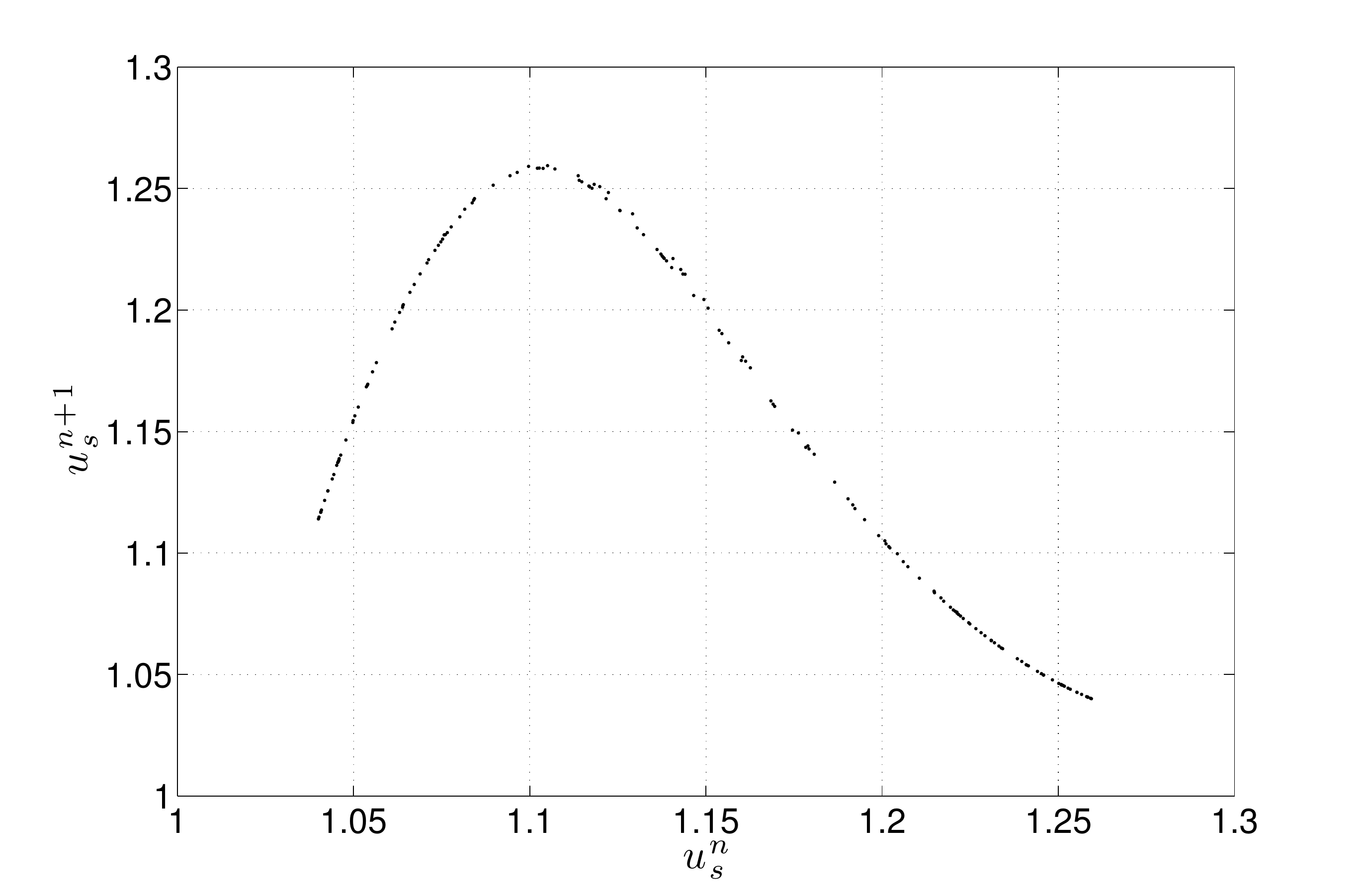}
\par\end{raggedright}

\caption{\label{fig:Lorenz-map}The Lorenz map showing consecutive local
maxima ($u_{s}^{n},\,\, u_{s}^{n+1}$) of the shock strength $u_{s}\left(t\right)$
over large times (from $t=3000$ to $t=6000$) for the chaotic case
at $\alpha=5.1$.}
\end{figure}

We plot the chaotic attractor at $\alpha=5.1$ in the space of $u_{s}$,
$\dot{u}_{s}$, and $\ddot{u}_{s}$ as shown in Fig. \ref{fig:Phase-plane}.
Its resemblance to the R\"ossler attractor \cite{Rossler1976} is evident.
Interestingly, when we plot the local maxima of $u_{s}$ versus their
prior values (i.e. the Lorenz map \cite{Lorenz1963}, see Fig. \ref{fig:Lorenz-map}),
the data fall almost on a curve. The curve also resembles the one
for the R\"ossler attractor. These observations suggest that the shock-wave
chaos arising from (\ref{eq:KFR-equation}) is controlled by a low-dimensional
process similar to that of a simple one-dimensional map$\--$just
as it is the case with the Lorenz and R\"ossler attractors \cite{strogatz1994nonlinear}. 
\begin{acknowledgments}
AK and LF gratefully acknowledge the support of KAUST. The work of
RRR was partially supported by the NSF grants DMS-1007967 and DMS-1115278.
\end{acknowledgments}
\bibliographystyle{plain}
%\bibliography{/Users/aslankasimov/Dropbox/Biblioteka/akasimov-refs}

\end{document}